\documentclass{article}%
\usepackage{t1enc}
\usepackage{amsmath}
\usepackage{sw20rui}%
\usepackage{amsfonts}%
\usepackage{amssymb}%
\usepackage{graphicx}
\begin{document}

\title{(Conformal) Killing Vectors and their Associated Bivectors}
\author{Garry Ludwig\\Department of Mathematical Sciences\\University of Alberta\\Edmonton, Alberta\\Canada T6G 2G1\\E-mail:gludwig@gpu.srv.ualberta.ca}
\date{February 13, 2002}
\maketitle
\tableofcontents

\begin{abstract}
Fayos and Sopuerta have recently set up a formalism for studying vacuum
spacetimes with an isometry, a formalism that is centred around the bivector
corresponding to the Killing vector and that adapts the tetrad to the
bivector. Steele has generalized their approach to include the homothetic
case. Here, we generalize this formalism to arbitrary spacetimes and to
homothetic and conformal Killing vectors but do not insist on aligning the
tetrad with the bivector. The most efficient way to use the formalism to find
conformal Killing vectors (proper or not) of a given spacetime is to combine
it with the notion of a preferred tetrad. A metric by Kimura is used as an
illustrative example.

\end{abstract}

\section{INTRODUCTION}

In a number of papers\cite{EL1,EL2,EL3} Brian Edgar and the present author
investigated spacetimes with (conformal) Killing vectors ((C)KVs), i.e. with
Killing vectors (KVs), a homothetic vector (HV), and/or proper conformal
Killing vectors (CKVs). One of the key ingredients was the notion of a
\textit{preferred} \textit{tetrad} relative to a (C)KV. Working, at first, in
the Geroch-Held-Penrose (GHP) formalism\cite{GHP} the notion of
\textit{preferred null directions} relative to a vector, in particular,
relative to a (C)KV, was defined. This was done in such a way that when a
suitably defined GHP generalization of the ordinary Lie derivative is applied
to such a preferred GHP tetrad the result takes its simplest possible form.
Proceeding to the Newman-Penrose formalism\cite{NP} it was then necessary to
define the notion of a \textit{preferred gauge} as well (thus defining what is
meant by a preferred tetrad). Although this was done by requiring that the the
GHP Lie operator and the ordinary Lie derivative have the same effect on
arbitrary scalar quantities, the upshot of it was that when the (ordinary) Lie
derivative is applied to a preferred tetrad it yields the simplest possible
result. In particular, relative to a KV, the Lie derivative annihilates the
tetrad if and only if the latter is preferred.

In a recent article, Fayos and Sopuerta\cite{FS} (FS) set up a formalism to
facilitate the study\ of vacuum spacetimes with an isometry. (Steele\cite{S}
has meanwhile extended their method to include homotheties.) This formalism
centred around the bivector associated with a Killing vector. The aim of the
present paper is three-fold. Firstly, we show that their formalism can be
obtained quite simply by re-writing the Killing equations and their
integrability conditions as obtained in Ref.[\cite{KL}] (KL), in terms of the
associated bivector. Secondly, since the latter equations were obtained for an
arbitrary spacetime, vacuum or not, and for homothetic and proper conformal
Killing vectors as well, the extension of the FS equations to this most
general case is straightforward. Thirdly, we show how the generalized FS
equations or, equivalently, the KL equations, can be used most efficiently to
find all (C)KVs for a given spacetime if they are combined with the notion of
a preferred tetrad\cite{EL2,EL3}.

Normally, to find all possible (C)KVs of a given spacetime one solves the
(conformal) Killing equations. Inevitably, this has to be done with the aid of
the integrability conditions of these equations. Both sets of equations were
worked out in all generality in KL\cite{KL} in the GHP formalism. They are
readily converted into the NP formalism. Generally speaking, these equations
are still quite difficult to tackle unless one chooses the tetrad
appropriately. Insisting that the tetrad be preferred relative to a (C)KV yet
to be found, furnishes tremendous simplification. In the FS approach the
tetrad direction(s) are chosen to be\ principal null direction(s) of the
bivector associated with the KV. In the case of an HV or a KV, such null
directions are then preferred and lead to suitable simplifications. (The issue
of alignment of the bivector with the Weyl tensor is not addressed here. It
applies only to some specialized cases, albeit perhaps interesting ones. Here
we concentrate on the more general case where such an alignment may or may not
exist.) However, it is usually better to adapt the null directions to the Weyl
tensor or some other aspect of the (conformal) geometry since they will then
be preferred with respect to \emph{all} (C)KVs. Except when there is
alignment, such null directions are then not principal null directions of the
bivector(s) associated with the (C)KVs. Further, when dealing with a proper
CKV, the principal null directions of the associated bivector are not
preferred and the equations will not simplify. Therefore, although the FS
formalism (and, by implication, the present extension) appears to be useful in
deriving general properties for spacetimes with (C)KVs it does not seem to be
a good tool for actually determining such (C)KVs unless combined with the
notion of a preferred tetrad. Although we maintain that the best tool is the
commutator approach\cite{EL1}$^{\text{-}}$\cite{EL3} (which also employs
preferred tetrads), in this paper we shall work with the KL equations or,
equivalently, the (generalized) FS equations.

The notation used here for the tetrad components of the (C)KV and of other
quantities agrees with that of Refs.[\cite{EL2,EL3}] and is different in some
respects from that used by FS\cite{FS}. However, the correspondence is readily made.

In section 2 we review the notion of preferred null directions relative to a
(C)KV and rewrite the conformal Killing equations and their integrabilty
conditions in terms of the bivector associated with a (C)KV. In the following
section we make the connection to the FS formalism and discuss a few general
results that may readily be obtained from this formalism. In section 4, after
reviewing the notion of a preferred gauge, we convert these equations from the
GHP formalism to the NP formalism. Finally, in section 5, we illustrate on a
concrete example (the non-vacuum metric of Kimura\cite{KM}) how these
equations, when used in conjunction with a preferred tetrad, can be solved to
yield all (C)KVs of the given metric.

\section{THE GHP CONFORMAL KILLING EQUATIONS AND THEIR INTEGRABILITY CONDITIONS}

The most useful generalization of the Lie derivative $\pounds_{_{\xi}}$ to the
GHP formalism is the ''GHP Lie derivative \L$_{_{\xi}}$'' defined as
\begin{equation}
\text{\L}_{_{\xi}}=L_{_{\xi}}-\frac{p}{4}\left(  \mathcal{P}-\mathcal{P}%
^{\prime}+\mathcal{P}^{\ast}-\mathcal{P}^{\prime\ast}\right)  -\frac{q}%
{4}\left(  \mathcal{P}-\mathcal{P}^{\prime}+\mathcal{P}^{\prime\ast
}-\mathcal{P}^{\ast}\right) \label{A1}%
\end{equation}
where $p,q$ are the GHP weights of the quantity operated upon,%
\begin{equation}
\mathcal{P}=n_{_{\mu}}L_{_{\xi}}l^{^{\mu}}\label{A2}%
\end{equation}
(with similar definitions for the companions\cite{KL} $\mathcal{P}^{\prime
},\mathcal{P}^{\ast},\mathcal{P}^{\prime\ast}$), and
\begin{equation}
L_{_{\xi}}=\pounds_{_{\xi}}-\xi^{^{\mu}}\left(  p\zeta_{_{\mu}}+q\overline
{\zeta}_{_{\mu}}\right)  ,\label{A3}%
\end{equation}
where, using the usual NP notation,%
\begin{equation}
\zeta_{_{\mu}}=\gamma l_{_{\mu}}+\varepsilon n_{_{\mu}}-\alpha m_{_{\mu}%
}-\beta\overline{m}_{_{\mu}}.\label{A4}%
\end{equation}
As a result, \L$_{_{\xi}}l^{\mu}$ has the form%
\begin{equation}
\text{\L}_{_{\xi}}l^{^{\mu}}=\frac{1}{2}\left(  \mathcal{P}+\mathcal{P}%
^{\prime}\right)  l^{^{\mu}}+\mathcal{R}n^{^{\mu}}-\mathcal{Q}m^{^{\mu}%
}-\overline{\mathcal{Q}}\overline{m}^{^{\mu}}\label{A5}%
\end{equation}
and similarly for its companions under the prime, star and star-prime
operations. When $\xi$ is a (C)KV, i.e. when it satisfies%
\begin{equation}
\pounds_{_{\xi}}g_{\mu\nu}=\varphi g_{\mu\nu},\label{A6}%
\end{equation}
the conformal Killing equations may be written
\begin{align}
\mathcal{P}^{\prime}  & =-\mathcal{P}-\varphi\text{, \hspace{0.6cm}%
}\mathcal{P}^{\prime\ast}=-\mathcal{P}^{\ast}-\varphi\nonumber\\
\mathcal{Q}  & =\mathcal{Q}^{\prime\ast}\text{, \hspace{0.6cm}}\mathcal{Q}%
^{\prime}=\mathcal{Q}^{\ast}\nonumber\\
\mathcal{R}  & =\mathcal{R}^{\prime}=\mathcal{R}^{\ast}=\mathcal{R}%
^{\prime\ast}=0.\label{A7}%
\end{align}
Since\cite{L} $\overline{\mathcal{P}^{\ast}}=\mathcal{P}^{\prime\ast}$ it
follows from the second of Eqs.(\ref{A7}) that $\overline{\mathcal{P}^{\ast}%
}+\mathcal{P}^{\ast}=-\varphi$. Hence, for a (C)KV, we can write%
\begin{equation}
\mathcal{P}^{\ast}=-\frac{\varphi}{2}-i\mathcal{S}\label{A7a}%
\end{equation}
where $S$ is real. It follows that for a (C)KV,
\begin{equation}
\text{\L}_{_{\xi}}=L_{_{\xi}}-\frac{p}{2}\left(  \mathcal{P}-i\mathcal{S}%
+\frac{\varphi}{2}\right)  -\frac{q}{2}\left(  \mathcal{P}+i\mathcal{S}%
+\frac{\varphi}{2}\right)  .\label{A7b}%
\end{equation}

When the null directions $l$ and $n$ are chosen such that $\mathcal{Q}%
=\mathcal{Q}^{\prime}=0$, which is always possible, they are called
\emph{preferred}. In this case we have (for a (C)KV $\xi$)
\begin{equation}
\text{\L}_{_{\xi}}l^{^{\mu}}=-\frac{1}{2}\varphi l^{^{\mu}},\label{A8}%
\end{equation}
and similarly for its companions.

The bivector $F_{\mu\nu}$ associated with a (C)KV is, as in Eq.(KL24) of
Ref.[\cite{KL}], defined by%
\begin{equation}
F_{\mu\nu}=\xi_{\mu;\nu}-\frac{1}{2}\varphi g_{\mu\nu}\label{A9}%
\end{equation}
and its tetrad components, as in Eqs.(KL26-KL28), are given by%
\begin{align}
\phi_{_{0}}  & =F_{\mu\nu}l^{^{\mu}}m^{^{\nu}}=\overline{\mathcal{Q}}-\kappa
a-\tau b+\sigma c+\rho\overline{c}\label{A10}\\
\phi_{_{2}}  & =F_{\mu\nu}\overline{m}^{^{\mu}}n^{^{\nu}}=-\overline
{\mathcal{Q}}^{\prime}-\pi a-\nu b+\mu c+\lambda\overline{c}\tag{12$^\prime
$}\\
\phi_{_{1}}  & =\frac{1}{2}F_{\mu\nu}\left(  l^{^{\mu}}n^{^{\nu}}+\overline
{m}^{^{\mu}}m^{^{\nu}}\right)  =\frac{1}{2}\left(  \mathcal{P}-i\mathcal{S}%
+\frac{\varphi}{2}\right)  ,\label{A10a}%
\end{align}
where $a$, $b$, $c$, $d$ are the tetrad components of the (C)KV $\xi$:%
\begin{equation}
\xi=al^{^{\mu}}+bn^{^{\mu}}-cm^{^{\mu}}-\overline{c}\overline{m}^{^{\mu}%
}.\label{A11}%
\end{equation}
Note that by the last of Eqs.(\ref{A10}), Eq.(\ref{A7b}) may now be written as%
\begin{equation}
\text{\L}_{_{\xi}}=L_{_{\xi}}-\frac{p}{2}\phi_{_{1}}-\frac{q}{2}\overline
{\phi_{_{1}}}.\label{A11b}%
\end{equation}
It is also worthwhile noting that the weights of $\phi_{_{0}}$, $\phi_{_{1}}$,
$\phi_{_{2}}$ are, respectively, (2,0), (0,0), and \linebreak (-2,0) and that
under the prime and star operations\cite{KL,L} these quantities transform
according to
\begin{align}
\phi_{_{0}}^{\prime}  & =-\phi_{_{2}},\qquad\phi_{_{1}}^{\prime}=-\phi_{_{1}%
}^{\prime}\nonumber\\
\phi_{_{i}}^{\ast}  & =\phi_{_{i}}\quad{\small (i=0,1,2).}\label{A11a}%
\end{align}
The weights and transformation properties of other GHP quantities of interest
are found in the Appendices of Ref.[\cite{KL}].

In terms of these quantities $\phi_{i}$ the Killing equations as given by
Eqs.(KL21-KL23) become%
\begin{align}
\text{\th}b  & =-\kappa c-\overline{\kappa}\overline{c}\label{A12}\\
\text{\th}^{\prime}a  & =\overline{\nu}c+\nu\overline{c}\tag{17$^\prime$}\\
\text{\dh}\overline{c}  & =-\sigma a+\overline{\lambda}b\tag{17*}\\
& \nonumber\\
\text{\th}a  & =\frac{\varphi}{2}-\phi_{_{1}}-\overline{\phi_{_{1}}}%
+\overline{\pi}c+\pi\overline{c}\label{A13}\\
\text{\th}^{\prime}b  & =\frac{\varphi}{2}+\phi_{_{1}}+\overline{\phi_{_{1}}%
}-\tau c-\overline{\tau}\overline{c}\tag{18$^\prime$}\\
\text{\dh}c  & =-\frac{\varphi}{2}+\phi_{_{1}}-\overline{\phi_{_{1}}%
}-\overline{\rho}a+\mu b\tag{18*}\\
& \nonumber\\
\text{\th}c  & =-\overline{\phi_{_{0}}}-\overline{\kappa}a+\pi b\label{A14}\\
\text{\th}^{\prime}\overline{c}  & =\overline{\phi_{_{2}}}-\tau a+\overline
{\nu}b\tag{19$^\prime$}\\
\text{\dh}a  & =-\overline{\phi_{_{2}}}+\overline{\lambda}c+\mu\overline
{c}\tag{19*}\\
\overline{\text{\dh}}b  & =\overline{\phi_{_{0}}}-\rho c-\overline{\sigma
}\overline{c}\tag{19$^\prime$*}%
\end{align}
Their first integrability conditions, given in Eqs.(KL34-KL36) are also easily
re-written in terms of the $\phi_{i}$ and become%
\begin{align}
\text{\th}\phi_{_{1}}  & =\pi\phi_{_{0}}-\kappa\phi_{_{2}}-\frac{1}%
{4}\text{\th}\varphi+b\left(  \Lambda-\Phi_{_{11}}-\Psi_{_{2}}\right)
+c\Psi_{_{1}}+\overline{c}\Phi_{_{10}}\label{A15}\\
\text{\th}^{\prime}\phi_{_{1}}  & =\nu\phi_{_{0}}-\tau\phi_{_{2}}+\frac{1}%
{4}\text{\th}^{\prime}\varphi-a\left(  \Lambda-\Phi_{_{11}}-\Psi_{_{2}%
}\right)  -c\Phi_{_{12}}-\overline{c}\Psi_{_{3}}\tag{20$^\prime$}\\
\text{\dh}\phi_{_{1}}  & =\mu\phi_{_{0}}-\sigma\phi_{_{2}}-\frac{1}%
{4}\text{\dh}\varphi+a\Psi_{_{1}}-b\Phi_{_{12}}+\overline{c}\left(
\Lambda+\Phi_{_{11}}-\Psi_{_{2}}\right) \tag{20*}\\
\overline{\text{\dh}}\phi_{_{1}}  & =\lambda\phi_{_{0}}-\rho\phi_{_{2}}%
+\frac{1}{4}\overline{\text{\dh}}\varphi+a\Phi_{_{10}}-b\Psi_{_{3}}-c\left(
\Lambda+\Phi_{_{11}}-\Psi_{_{2}}\right) \tag{20$^\prime$*}\\
& \nonumber\\
\text{\th}\phi_{_{2}}  & =2\pi\phi_{_{1}}-\frac{1}{2}\overline{\text{\dh}%
}\varphi-b\left(  \Psi_{_{3}}+\Phi_{_{21}}\right)  +c\left(  \Psi_{_{2}%
}+2\Lambda\right)  +\overline{c}\Phi_{_{20}}\label{A16}\\
\text{\th}^{\prime}\phi_{_{0}}  & =-2\tau\phi_{_{1}}+\frac{1}{2}\text{\dh
}\varphi+a\left(  \Psi_{_{1}}+\Phi_{_{01}}\right)  -c\Phi_{_{02}}-\overline
{c}\left(  \Psi_{_{2}}+2\Lambda\right) \tag{21$^\prime$}\\
\text{\dh}\phi_{_{2}}  & =2\mu\phi_{_{1}}-\frac{1}{2}\text{\th}^{\prime
}\varphi+a\left(  \Psi_{_{2}}+2\Lambda\right)  -b\Phi_{_{22}}+\overline
{c}\left(  \Phi_{_{21}}-\Psi_{_{3}}\right) \tag{21*}\\
\overline{\text{\dh}}\phi_{_{0}}  & =-2\rho\phi_{_{1}}+\frac{1}{2}\text{\th
}\varphi+a\Phi_{_{00}}-b\left(  \Psi_{_{2}}+2\Lambda\right)  +c\left(
\Psi_{_{1}}-\Phi_{_{01}}\right) \tag{21$^\prime$*}\\
& \nonumber\\
\text{\th}\phi_{_{0}}  & =-2\kappa\phi_{_{1}}-b\left(  \Psi_{_{1}}+\Phi
_{_{01}}\right)  +c\Psi_{_{0}}+\overline{c}\Phi_{_{00}}\label{A17}\\
\text{\th}^{\prime}\phi_{_{2}}  & =2\nu\phi_{_{1}}+a\left(  \Psi_{_{3}}%
+\Phi_{_{21}}\right)  -c\Phi_{_{22}}-\overline{c}\Psi_{_{4}}\tag{22$^\prime
$}\\
\text{\dh}\phi_{_{0}}  & =-2\sigma\phi_{_{1}}+a\Psi_{_{0}}-b\Phi_{_{02}%
}-\overline{c}\left(  \Psi_{_{1}}-\Phi_{_{01}}\right) \tag{22*}\\
\overline{\text{\dh}}\phi_{_{2}}  & =2\lambda\phi_{1}+a\Phi_{_{20}}%
-b\Psi_{_{4}}+c\left(  \Psi_{_{3}}-\Phi_{_{21}}\right) \tag{22$^\prime$*}%
\end{align}
The Maxwell equations (with source) are implicit in these equations. They are
obtained by subtracting Eq.(\ref{A16}$^{\prime\ast}$) from Eq.(\ref{A15}) and
doing the same with their respective companions. Thus,
\begin{align}
\text{\th}\phi_{_{1}}-\overline{\text{\dh}}\phi_{_{0}}  & =\pi\phi_{_{0}%
}+2\rho\phi_{_{1}}-\kappa\phi_{_{2}}-\frac{3}{4}\text{\th}\varphi-a\Phi
_{_{00}}+b\left(  3\Lambda-\Phi_{_{11}}\right)  +c\Phi_{_{01}}+\overline
{c}\Phi_{_{10}}\label{A18}\\
\text{\th}^{\prime}\phi_{_{1}}-\text{\dh}\phi_{_{2}}  & =\nu\phi_{_{0}}%
-2\mu\phi_{_{1}}-\tau\phi_{_{2}}+\frac{3}{4}\text{\th}^{\prime}\varphi
-a\left(  3\Lambda-\Phi_{_{11}}\right)  +b\Phi_{_{22}}-c\Phi_{_{12}}%
-\overline{c}\Phi_{_{21}}\tag{23$^\prime$}\\
\text{\dh}\phi_{_{1}}-\text{\th}^{\prime}\phi_{_{0}}  & =\mu\phi_{_{0}}%
+2\tau\phi_{_{1}}-\sigma\phi_{_{2}}-\frac{3}{4}\text{\dh}\varphi-a\Phi_{_{01}%
}-b\Phi_{_{12}}+c\Phi_{_{02}}+\overline{c}\left(  3\Lambda+\Phi_{_{11}}\right)
\tag{23*}\\
\overline{\text{\dh}}\phi_{_{1}}-\text{\th}\phi_{_{2}}  & =\lambda\phi_{_{0}%
}-2\pi\phi_{_{1}}-\rho\phi_{_{2}}+\frac{3}{4}\overline{\text{\dh}}%
\varphi+a\Phi_{_{10}}+b\Phi_{_{21}}-\overline{c}\Phi_{_{20}}-c\left(
3\Lambda+\Phi_{_{11}}\right) \tag{23$^\prime$*}%
\end{align}
When specialized to vacuum and to a proper Killing vector, Eqs.(\ref{A15}) -
(\ref{A17}) readily yield the formalism of FS, as we shall see in the next section.

\section{GENERAL CONSIDERATIONS}

Before revisiting the FS equations, albeit in their most general from which
includes non-vacuum metrics and HV and proper CKV, let us derive some general
results of interest.

Multiplying Eqs.(\ref{A17}), (\ref{A16}$^{\prime})$, (\ref{A17}*), and
(\ref{A16}$^{\prime}$*) by $a,b,-c,-\overline{c}$, respectively, and adding
the results we first calculate $\left(  a\text{\th}+b\text{\th}^{\prime
}-c\text{\dh}-\overline{c}\overline{\text{\dh}}\right)  \phi_{_{0}}$, i.e.
$L_{_{\xi}}\phi_{_{0}}$, and then, from Eq.(\ref{A11b}), \L$_{_{\xi}}%
\phi_{_{0}}$. Doing similar calculations for $\phi_{_{1}}$ and $\phi_{_{2}}$
we find that
\begin{align}
\text{\L}_{_{\xi}}\phi_{_{0}}  & =-2\overline{\mathcal{Q}}\phi_{_{1}}+\frac
{1}{2}b\text{\dh}\varphi-\frac{1}{2}\overline{c}\text{\th}\varphi\label{B1}\\
\text{\L}_{_{\xi}}\phi_{_{2}}  & =-2\overline{\mathcal{Q}}^{\prime}\phi_{_{1}%
}-\frac{1}{2}a\overline{\text{\dh}}\varphi+\frac{1}{2}c\text{\th}^{\prime
}\varphi\tag{24$^\prime$}\\
\text{\L}_{_{\xi}}\phi_{_{1}}  & =-\overline{\mathcal{Q}}^{\prime}\phi_{_{0}%
}-\overline{\mathcal{Q}}\phi_{_{2}}+\frac{1}{4}\left(  -a\text{\th}%
\varphi+b\text{\th}^{\prime}\varphi+c\text{\dh}\varphi-\overline{c}%
\overline{\text{\dh}}\varphi\right)  .\label{B2}%
\end{align}

Restricting ourselves to HV and KVs and assuming that the bivector $F_{\mu\nu
}$ does not vanish identically, we see immediately from Eqs.(\ref{B1},
24$^{\prime}$, \ref{B2}) that if the null directions are preferred then the
$\phi_{_{i}}$ are annihilated by the GHP operator \L$_{_{\xi}}$, as perhaps
expected. Conversely, if we choose the $l$ - direction to be a principal null
direction of the bivector, so that $\phi_{_{0}}$vanishes, we find immediately
from Eqs.(\ref{B1}) and (\ref{B2}) that $\mathcal{Q}=0$, i.e. that this
direction is preferred. From the first of Eqs.(\ref{A10}) we now deduce that
\begin{equation}
\kappa a+\tau b-\sigma c-\rho\overline{c}=0\label{B3}%
\end{equation}
i.e. that the vector $X_{_{1}}=-\tau l-\kappa n+\rho m+\sigma\overline{m}$ is
orthogonal to the KV/HV $\xi$. If the bivector is non-null we can choose the
second null direction to coincide with the bivector's second principal null
direction and we get the prime of the above result, $\mathcal{Q}^{\prime}=0$,
i.e.
\begin{equation}
-\pi a-b\nu+\mu c+\lambda\overline{c}=0\label{B4}%
\end{equation}
i.e. that the vector $X_{_{3}}=\nu l+\pi n-\lambda m-\mu\overline{m}$ is also
orthogonal to the KV/HV $\xi$. As FS have shown, and as we shall see below,
there may, under certain circumstances (such as in vacuum), be yet another
vector $X_{_{2}}$ orthogonal to $\xi$.

Still restricting ourselves to an HV or a KV, if we add Eqs.(\ref{A16}*) and
(\ref{A16}$^{\prime}$*), convert to the NP formalism where we take a gauge
such that $\rho=\mu$ (which is possible provided $\rho\mu\neq0$) we obtain
after a lengthy calculation that $\rho,_{_{\mu}}\xi^{_{\mu}}=\frac{\varphi}%
{2}\rho$. Clearly now, if $\rho$ is a constant in this gauge, $\varphi$ has to
vanish; there cannot be an HV. This is the case for the Kimura metric
considered in Section 5. It should be noted that this conclusion is arrived at
much faster using the KL formalism\cite{KL}. In fact, Eqs.(\ref{A16}*) and
(\ref{A16}$^{\prime}$*) are simply Eqs.(KL35*) and (KL35$^{\prime}$*) which,
for preferred null directions reduce to the complex conjugates of \L$_{_{\xi}%
}\rho=\frac{\varphi}{2}\rho$ and \L$_{_{\xi}}\mu=\frac{\varphi}{2}\mu$,
respectively. Since the gauge $\rho=\mu$ has a preferred boost part and since
both $\rho$ and $\mu$ have weights of the form ($p,p$), it follows from the
discussion in the next section that $\pounds_{_{\xi}}\rho=\frac{\varphi}%
{2}\rho$ and $\pounds_{_{\xi}}\mu=\frac{\varphi}{2}\mu$. Hence, when $\rho
=\mu=$ constant we necessarily have $\varphi=0$.

Let us now return to the general case that includes proper CKVs.
Eqs.(\ref{A16})-(\ref{A17}), including their companions, can be solved
pairwise for $N\Psi_{_{i}}$ $\left(  i=0,...4)\right)  $, where $N=\xi^{^{\mu
}}\xi_{_{\mu}}$. We find that
\begin{align}
N\Psi_{_{3}}  & =c\left(  \text{\dh}\phi_{_{2}}-2\mu\phi_{_{1}}+\frac{1}%
{2}\text{\th}^{\prime}\varphi+b\Phi_{_{22}}-\overline{c}\Phi_{_{21}}\right)
\nonumber\\
& -a\left(  \text{\th}\phi_{_{2}}-2\pi\phi_{_{1}}+\frac{1}{2}\overline
{\text{\dh}}\varphi+b\Phi_{_{21}}-\overline{c}\Phi_{_{20}}\right)
\label{B5a}\\
N\left(  \Psi_{_{2}}+2\Lambda\right)   & =b\left(  \text{\dh}\phi_{_{2}}%
-2\mu\phi_{_{1}}+\frac{1}{2}\text{\th}^{\prime}\varphi+b\Phi_{_{22}}%
-\overline{c}\Phi_{_{21}}\right) \nonumber\\
& -\overline{c}\left(  \text{\th}\phi_{_{2}}-2\pi\phi_{_{1}}+\frac{1}%
{2}\overline{\text{\dh}}\varphi+b\Phi_{_{21}}-\overline{c}\Phi_{_{20}}\right)
\label{B6}%
\end{align}%

\begin{align}
N\Psi_{_{1}}  & =b\left(  \text{\th}^{\prime}\phi_{_{0}}+2\tau\phi_{_{1}%
}-\frac{1}{2}\text{\dh}\varphi-a\Phi_{_{01}}+c\Phi_{_{02}}\right) \nonumber\\
& -\overline{c}\left(  \overline{\text{\dh}}\phi_{_{0}}+2\rho\phi_{_{1}}%
-\frac{1}{2}\text{\th}\varphi-a\Phi_{_{00}}+c\Phi_{_{01}}\right)
\tag{28$^\prime$}\\
N\left(  \Psi_{_{2}}+2\Lambda\right)   & =c\left(  \text{\th}^{\prime}%
\phi_{_{0}}+2\tau\phi_{_{1}}-\frac{1}{2}\text{\dh}\varphi-a\Phi_{_{01}}%
+c\Phi_{_{02}}\right) \nonumber\\
& -a\left(  \overline{\text{\dh}}\phi_{_{0}}+2\rho\phi_{_{1}}-\frac{1}%
{2}\text{\th}\varphi-a\Phi_{_{00}}+c\Phi_{_{01}}\right) \tag{29$^\prime$}%
\end{align}%
\begin{align}
N\Psi_{_{0}}  & =b\left(  \text{\dh}\phi_{_{0}}+2\sigma\phi_{_{1}}%
+b\Phi_{_{02}}-\overline{c}\Phi_{_{01}}\right)  -\overline{c}\left(
\text{\th}\phi_{_{0}}+2\kappa\phi_{_{1}}+b\Phi_{_{01}}-\overline{c}\Phi
_{_{00}}\right) \label{B7}\\
N\Psi_{_{1}}  & =-a\left(  \text{\th}\phi_{_{0}}+2\kappa\phi_{_{1}}%
+b\Phi_{_{01}}-\overline{c}\Phi_{_{00}}\right)  +c\left(  \text{\dh}\phi
_{_{0}}+2\sigma\phi_{_{1}}+b\Phi_{_{02}}-\overline{c}\Phi_{_{01}}\right)
\label{B8}%
\end{align}%
\begin{align}
N\Psi_{_{4}}  & =c\left(  \text{\th}^{\prime}\phi_{_{2}}-2\nu\phi_{_{1}}%
-a\Phi_{_{21}}+c\Phi_{_{22}}\right)  -a\left(  \overline{\text{\dh}}\phi
_{_{2}}-2\lambda\phi_{_{1}}-a\Phi_{_{20}}+c\Phi_{_{21}}\right) \tag{30$%
^\prime$}\\
N\Psi_{_{3}}  & =b\left(  \text{\th}^{\prime}\phi_{_{2}}-2\nu\phi_{_{1}}%
-a\Phi_{_{21}}+c\Phi_{_{22}}\right)  -\overline{c}\left(  \overline{\text{\dh
}}\phi_{_{2}}-2\lambda\phi_{_{1}}-a\Phi_{_{20}}+c\Phi_{_{21}}\right)
\tag{31$^\prime$}%
\end{align}
Note that these equations hold even when $N=0$. Together with the Maxwell
equations (\ref{A18}) they are equivalent to the first integrability
conditions we started with. For vacuum and when $\xi$ is a Killing vector,
they reduce to those of the FS formalism, provided we assume either that the
bivector is nonnull and $\phi_{_{0}}=\phi_{_{2}}=0$, $\phi_{_{1}}\neq0$ or
that the bivector is null and $\phi_{_{0}}=\phi_{_{1}}=0$, $\phi_{_{2}}\neq0$.
The latter formalism is indeed a very special case of the present one.

Let us assume that $\xi$ is a KV with a nonsingular bivector and that we have
taken a canonical basis for which $\phi_{_{0}}=0=\phi_{_{2}}.$ Adding
Eqs.(\ref{B6}) and (\ref{B6}$^{\prime}$) then yields%
\begin{equation}
2\phi_{_{1}}\xi\cdot X_{2}-b^{2}\Phi_{_{22}}+2b\overline{c}\Phi_{_{21}%
}-c\overline{c}\Phi_{_{20}}-2ac\Phi_{_{01}}+c^{2}\Phi_{_{02}}+a^{2}\Phi
_{_{00}}=0\label{B9}%
\end{equation}
where%
\begin{equation}
X_{2}=\mu l-\rho n-\pi m+\tau\overline{m}.\label{B10}%
\end{equation}
In vacuum this gives the third vector orthogonal to $\xi$, as also derived by
FS. But we see clearly from Eq.(\ref{B9}) that only under special conditions
as the ones described do we get such a third orthogonal vector.

\ \ \ 

\section{THE NP CONFORMAL KILLING EQUATIONS AND THEIR INTEGRABILITY CONDITIONS}

From Eqs.(\ref{A1})-(\ref{A4}) we see that
\begin{equation}
\text{\L}_{_{\xi}}=\pounds_{_{\xi}}-p\mathfrak{G}-q\overline{\mathfrak{G}%
}\label{C1}%
\end{equation}
where
\begin{equation}
\mathfrak{G}=\frac{1}{4}\left(  \mathcal{P}-\mathcal{P}^{\prime}%
+\mathcal{P}^{\ast}-\mathcal{P}^{\prime\ast}+4\xi^{^{\mu}}\zeta_{_{\mu}%
}.\right) \label{C2}%
\end{equation}
Because of the conformal Killing equations (\ref{A7}), for a (C)KV $\xi$ this
reduces to%
\begin{equation}
\mathfrak{G}=\frac{1}{2}\left(  \mathcal{P}-i\mathcal{S}+\frac{\varphi}%
{2}+2\xi^{^{\mu}}\zeta_{_{\mu}}.\right) \label{C3}%
\end{equation}
In terms of the component $\phi_{_{1}}$ of the associated bivector this
becomes
\begin{equation}
\mathfrak{G}=\phi_{_{1}}+\xi^{^{\mu}}\zeta_{_{\mu}},\label{C4}%
\end{equation}
i.e.
\begin{equation}
\mathfrak{G}=\phi_{_{1}}+\epsilon a+\gamma b-\beta c-\alpha\overline
{c}.\label{C5}%
\end{equation}

To fix the gauge requires the specification of two real parameters
corresponding to boost and phase. It is therefore possible to have a preferred
boost or a preferred phase (or both). The necessary and sufficient condition
for the boost-part of the gauge to be preferred is that \L$_{_{\xi}}\eta=$
$\pounds_{_{\xi}}\eta$ for any scalar quantity $\eta$ of weight ($p,p$).
Therefore, for a preferred boost we must have
\begin{equation}
\mathfrak{G}+\overline{\mathfrak{G}}=0.\label{C6}%
\end{equation}
Similarly, the necessary and sufficient condition for the phase-part of the
gauge to be preferred is that \L$_{_{\xi}}\eta=$ $\pounds_{_{\xi}}\eta$ for
any scalar quantity $\eta$ of weight ($p,-p$), i.e. that%
\begin{equation}
\mathfrak{G}-\overline{\mathfrak{G}}=0.\label{C7}%
\end{equation}

The necessary and sufficient condition for the full gauge to be preferred is
that $\mathfrak{G}$ vanish, i.e. that%
\begin{equation}
\phi_{_{1}}=-\epsilon a-\gamma b+\beta c+\alpha\overline{c}.\label{C8}%
\end{equation}

Recalling that%
\begin{equation}
\text{\th}=D-p\epsilon-q\overline{\epsilon},\label{C9}%
\end{equation}
and similarly for its companions, it is straightforward to write
Eqs.(\ref{A12})-(\ref{A17}) in NP notation. They become
\begin{align}
Db-\left(  \epsilon+\overline{\epsilon}\right)  b  & =-\kappa c-\overline
{\kappa}\overline{c}\label{C10}\\
Da+\left(  \epsilon+\overline{\epsilon}\right)  a  & =\frac{\varphi}{2}%
-\phi_{_{1}}-\overline{\phi_{_{1}}}+\overline{\pi}c+\pi\overline{c}%
\label{C11}\\
Dc+\left(  \epsilon-\overline{\epsilon}\right)  c  & =-\overline{\phi_{_{0}}%
}-\overline{\kappa}a+\pi b\label{C12}\\
D\phi_{_{1}}  & =\pi\phi_{_{0}}-\kappa\phi_{_{2}}-\frac{1}{4}D\varphi+b\left(
\Lambda-\Phi_{_{11}}-\Psi_{_{2}}\right)  +c\Psi_{_{1}}+\overline{c}\Phi
_{_{10}}\label{C13}\\
D\phi_{_{2}}+2\epsilon\phi_{_{2}}  & =2\pi\phi_{_{1}}-\frac{1}{2}%
\overline{\delta}\varphi-b\left(  \Psi_{_{3}}+\Phi_{_{21}}\right)  +c\left(
\Psi_{_{2}}+2\Lambda\right)  +\overline{c}\Phi_{_{20}}\label{C14}\\
D\phi_{_{0}}-2\epsilon\phi_{_{0}}  & =-2\kappa\phi_{_{1}}-b\left(  \Psi_{_{1}%
}+\Phi_{_{01}}\right)  +c\Psi_{_{0}}+\overline{c}\Phi_{_{00}}\label{C15}%
\end{align}
together with their companions.

\section{AN ILLUSTRATIVE EXAMPLE}

Although the commutator approach (Ref. \cite{EL1,EL2,EL3}) seems preferable to
a direct solving of the Killing equations and their first integrability
conditions we illustrate in this section how the latter approach is
facilitated by using preferred tetrads relative to the (C)KVs. To this end we
once again\cite{EL3} find all (C)KVs for the Kimura metric\cite{KM}.

The Kimura metric considered by Koutras and Skea\cite{KS} is given by
\begin{equation}
ds^{2}=\frac{r^{2}}{b_{_{0}}}dt^{2}-\frac{1}{r^{2}b_{_{0}}^{2}}dr^{2}%
-r^{2}d\theta^{2}-r^{2}\sin^{2}\theta d\phi^{2},\label{D1}%
\end{equation}
where $b_{_{0}}$ is a constant. It is of Petrov type D with a non-zero energy
momentum tensor. We can readily construct a tetrad such that follows that
\begin{equation}%
\begin{array}
[c]{llll}%
Dt=\frac{\sqrt{b_{_{_{0}}}}}{r\sqrt{2}}, & Dr=\frac{rb_{_{0}}}{\sqrt{2}}, &
D\theta=0, & D\phi=0\\
\triangle t=\frac{\sqrt{b_{_{_{0}}}}}{r\sqrt{2}}, & \triangle r=-\frac
{rb_{_{0}}}{\sqrt{2}}, & \triangle\theta=0, & \triangle\phi=0\\
\delta t=0, & \delta r=0, & \delta\theta=\frac{1}{r\sqrt{2}}, & \delta
\phi=\frac{i}{r\sqrt{2}\sin\theta}.
\end{array}
\label{D2}%
\end{equation}

The NP spin coefficients are then given by
\begin{align}
\kappa & =\sigma=\lambda=\nu=\tau=\pi=0\nonumber\\
\gamma & =\varepsilon=\frac{b_{_{_{0}}}}{2\sqrt{2}},\qquad\rho=\mu
=-\frac{b_{_{_{0}}}}{\sqrt{2}},\qquad\beta=-\overline{\alpha}=\frac{\cot
\theta}{2\sqrt{2}r},\label{D3}%
\end{align}
and
\begin{equation}
\Psi_{_{2}}=-\frac{1}{6r^{2}},\qquad\Phi_{_{11}}=\frac{1}{4r^{2}}%
,\qquad\Lambda=-\frac{b_{_{0}}^{2}}{2}+\frac{1}{12r^{2}}\label{D4}%
\end{equation}
are the only nonzero components of the Riemann tensor.

In view of Eqs.(\ref{D3}) and (\ref{D4}), the conformal Killing equations and
their first integrability conditions, Eqs. (\ref{C10}) -(\ref{C15}) and their
companions, become%
\begin{align}
Da  & =\frac{\varphi}{2}-\left(  \mathfrak{G}+\overline{\mathfrak{G}}\right)
+\frac{b_{_{_{0}}}}{\sqrt{2}}b,\qquad\quad\Delta a=-\frac{b_{_{_{0}}}}%
{\sqrt{2}}a,\quad\qquad\delta a=\overline{\delta}a=0\nonumber\\
Db  & =\frac{b_{_{_{0}}}}{\sqrt{2}}b,\qquad\quad\Delta b=\frac{\varphi}%
{2}+\mathfrak{G}+\overline{\mathfrak{G}}-\frac{b_{_{_{0}}}}{\sqrt{2}}%
a,\quad\qquad\delta b=\overline{\delta}b=0\nonumber\\
Dc  & =\frac{b_{_{_{0}}}}{\sqrt{2}}c,\quad\qquad\Delta c=-\frac{b_{_{_{0}}}%
}{\sqrt{2}}c,\nonumber\\
\delta c  & =-\frac{\varphi}{2}+\frac{b_{_{_{0}}}}{\sqrt{2}}\left(
a-b\right)  +\mathfrak{G}-\overline{\mathfrak{G}}-\frac{\cot\theta}{\sqrt{2}%
r}\overline{c},\quad\qquad\overline{\delta}c=\frac{\cot\theta}{\sqrt{2}%
r}c\label{D5}%
\end{align}%

\begin{align}
D\phi_{_{0}}  & =\frac{b_{_{_{0}}}}{\sqrt{2}}\phi_{_{0}},\quad\qquad\Delta
\phi_{_{0}}-\frac{b_{_{_{0}}}}{\sqrt{2}}\phi_{_{0}}=\frac{1}{2}\delta
\varphi+\overline{c}b_{_{0}}^{2}\nonumber\\
\delta\phi_{_{0}}  & =\frac{\cot\theta}{\sqrt{2}r}\phi_{_{0}},\quad
\qquad\overline{\delta}\phi_{_{0}}+\frac{\cot\theta}{\sqrt{2}r}\phi_{_{0}%
}=\sqrt{2}b_{_{_{0}}}+\frac{1}{2}D\varphi+bb_{_{0}}^{2}\label{D6}%
\end{align}%
\begin{align}
D\phi_{_{1}}  & =-\frac{1}{4}D\varphi-\frac{1}{2}bb_{_{0}}^{2},\quad
\qquad\Delta\phi_{1}=\frac{1}{4}\Delta\varphi+\frac{1}{2}ab_{_{0}}%
^{2}\nonumber\\
\delta\phi_{_{1}}  & =-\frac{b_{_{_{0}}}}{\sqrt{2}}\phi_{_{0}}-\frac{1}%
{4}\delta\varphi+\overline{c}\left(  \frac{1}{2r^{2}}-\frac{b_{_{0}}^{2}}%
{2}\right)  ,\quad\overline{\delta}\phi_{_{1}}=\frac{b_{_{_{0}}}}{\sqrt{2}%
}\phi_{_{2}}+\frac{1}{4}\overline{\delta}\varphi-c\left(  \frac{1}{2r^{2}%
}-\frac{b_{_{0}}^{2}}{2}\right) \label{D7}%
\end{align}%
\begin{align}
D\phi_{_{2}}+\frac{b_{_{_{0}}}}{\sqrt{2}}\phi_{_{2}}  & =-\frac{1}{2}%
\overline{\delta}\varphi-cb_{_{0}}^{2},\quad\qquad\Delta\phi_{_{2}}%
=-\frac{b_{_{_{0}}}}{\sqrt{2}}\phi_{_{2}}\nonumber\\
\delta\phi_{_{2}}+\frac{\cot\theta}{\sqrt{2}r}\phi_{_{2}}  & =-\sqrt
{2}b_{_{_{0}}}\phi_{_{1}}-\frac{1}{2}\Delta\varphi-ab_{_{0}}^{2},\quad
\qquad\overline{\delta}\phi_{_{2}}=\frac{\cot\theta}{\sqrt{2}r}\phi_{_{2}%
}.\label{D8}%
\end{align}

Clearly, the null directions are the principal null directions of the Weyl
tensor. These directions are preferred relative to all possible (C)KVs and
hence $\mathcal{Q}=$ $0$ $=\mathcal{Q}^{\prime}$. In view of Eqs.(\ref{D3}),
we see immediately from Eqs.(\ref{A10}) that%
\begin{equation}
\phi_{_{0}}=-\frac{b_{_{_{0}}}}{\sqrt{2}}\overline{c},\qquad\phi_{_{2}}%
=-\frac{b_{_{_{0}}}}{\sqrt{2}}c.\label{D9}%
\end{equation}
The gauge is not in any obvious way preferred for all (C)KVs; in fact, in
hindsight it will be seen that neither the boost-part nor the gauge part can
be chosen in a way that is preferred relative to all six (C)KVs that this
metric turns out to possess. Although we could solve the basic equations
involved for $\phi_{_{1}}$, it is easier to work in terms of $\mathfrak{G}$,
the quantity that is a measure of by how much the given gauge differs from a
preferred one for each (C)KV. Alternatively, we could put an arbitrary gauge
factor into our spin coefficients, but that too turns out to make the problem
more difficult. From Eqs.(\ref{C5}) and (\ref{D3}) we obtain%
\begin{equation}
\phi_{_{1}}=\mathfrak{G}-\frac{b_{_{0}}}{2\sqrt{2}}\left(  a+b\right)
+\frac{\cot\theta}{2\sqrt{2}r}\left(  c-\overline{c}\right)  .\label{D10}%
\end{equation}

Substituting from Eqs.(\ref{D9}) and (\ref{D10}) into Eqs. (\ref{D6}) and
(\ref{D8}) yields only that%
\begin{equation}
D\varphi=\sqrt{2}b_{_{_{0}}}\left(  \frac{\varphi}{2}-\mathfrak{G}%
-\overline{\mathfrak{G}}\right)  ,\quad\Delta\varphi=-\sqrt{2}b_{_{_{0}}%
}\left(  \frac{\varphi}{2}+\mathfrak{G}+\overline{\mathfrak{G}}\right)
,\quad\delta\varphi=0.\label{D11}%
\end{equation}
Subtracting these two equations we get $D\varphi-\Delta\varphi=\sqrt
{2}b_{_{_{0}}}\varphi$. From this we see immediately that the metric cannot
have a proper HV.

Substituting Eqs.(\ref{D9}) and (\ref{D10}) into Eqs.(\ref{D7}) and using the
Killing equations (\ref{D5}) yields
\begin{align}
D\mathfrak{G}  & =\Delta\mathfrak{G}=0\nonumber\\
\delta\mathfrak{G}  & =-\frac{\cot\theta}{2\sqrt{2}r}\left(  \mathfrak{G}%
-\overline{\mathfrak{G}}-\frac{\varphi}{2}+\frac{b_{_{_{0}}}\left(
a-b\right)  }{\sqrt{2}}\right)  +\frac{c+\overline{c}}{4r^{2}\sin^{2}\theta
}\nonumber\\
\overline{\delta}\mathfrak{G}  & \mathfrak{=}-\frac{\cot\theta}{2\sqrt{2}%
r}\left(  \mathfrak{G}-\overline{\mathfrak{G}}+\frac{\varphi}{2}%
-\frac{b_{_{_{0}}}\left(  a-b\right)  }{\sqrt{2}}\right)  -\frac
{c+\overline{c}}{4r^{2}\sin^{2}\theta}\label{D12}%
\end{align}
Eqs.(\ref{D5}), (\ref{D11}) and (\ref{D12}) can now be solved for the unknowns
$a,b,c,\varphi,\mathfrak{G}.$ We find that%
\begin{align}
a  & =r\left(  \frac{h_{_{0}}}{\sqrt{2b_{_{0}}}}+\frac{l_{_{1}}}{2\sqrt
{2}b_{_{0}}}+\frac{l_{_{0}}t}{\sqrt{2}}\right)  -\frac{l_{_{0}}}%
{\sqrt{2b_{_{0}}}}\nonumber\\
b  & =r\left(  \frac{h_{_{0}}}{\sqrt{2b_{_{0}}}}-\frac{l_{_{1}}}{2\sqrt
{2}b_{_{0}}}-\frac{l_{_{0}}t}{\sqrt{2}}\right)  -\frac{l_{_{0}}}%
{\sqrt{2b_{_{0}}}}\nonumber\\
c  & =\frac{-r}{\sqrt{2}}\left(  a_{_{0}}\cos\phi+b_{_{0}}\sin\phi\right)
+\frac{ir}{\sqrt{2}}\left[  c_{_{0}}\sin\theta-\cos\theta\left(  a_{_{0}}%
\sin\phi-b_{_{0}}\cos\phi\right)  \right] \nonumber\\
\varphi & =r\left(  l_{_{1}}+2l_{_{0}}b_{_{0}}t\right) \nonumber\\
\mathfrak{G}  & =-\frac{l_{_{0}}\sqrt{b_{_{0}}}}{2}+\frac{i}{2\sin\theta
}\left(  a_{_{0}}\sin\phi-b_{_{0}}\cos\phi\right) \label{D13}%
\end{align}
where $l_{_{0}},l_{_{1}},h_{_{0}},c_{_{0}},a_{_{0}},b_{_{0}}$ are integration
constants. Putting all but one of these equal to zero in turn and using
Eq.(\ref{A11}) we find, in coordinates ($t,r,\theta,\phi$), the two proper
CKVs
\begin{align}
l_{_{0}}  & =1:\qquad\xi_{_{\left(  1\right)  }}^{\mu}=\left(  -\frac{1}%
{r},r^{2}b_{_{0}}t,0,0\right) \nonumber\\
l_{_{1}}  & =1:\qquad\xi_{_{\left(  2\right)  }}^{\mu}=\left(  0,\frac{r^{2}%
}{2},0,0\right) \label{D14}%
\end{align}
with respective conformal factors $\varphi=2rb_{_{0}}t$ and $\varphi=r$, as
well as the four Killing vectors%
\begin{align}
h_{_{0}}  & =1:\qquad\xi_{_{\left(  3\right)  }}^{\mu}=\left(  1,0,0,0\right)
\nonumber\\
c_{_{0}}  & =1:\qquad\xi_{_{\left(  4\right)  }}^{\mu}=\left(  0,0,0,1\right)
\nonumber\\
a_{_{0}}  & =1:\qquad\xi_{_{\left(  5\right)  }}^{\mu}=\left(  0,0,\cos
\phi,-\cot\theta\sin\phi\right) \nonumber\\
b_{_{0}}  & =1:\qquad\xi_{_{\left(  6\right)  }}^{\mu}=\left(  0,0,\sin
\phi,\cot\theta\cos\phi\right)  .\label{D15}%
\end{align}

Noting from the last of Eqs.(\ref{D13}) that $\mathfrak{G}+\overline
{\mathfrak{G}}$ $=-l_{_{0}}\sqrt{b_{_{0}}}$ and $\mathfrak{G}-\overline
{\mathfrak{G}}=\frac{i}{\sin\theta}\left(  a_{_{0}}\sin\phi-b_{_{0}}\cos
\phi\right)  $ we see from Eqs.(\ref{D14}) and (\ref{D15}) that $\mathfrak{G}%
+\overline{\mathfrak{G}}$ vanishes for all but $\xi_{_{\left(  1\right)  }}.$
Therefore, the four KVs and the proper CKV $\xi_{_{\left(  2\right)  }}$ have
the boost-part of the given gauge preferred. Similarly, the two proper CKVs
and the KVs $\xi_{_{\left(  3\right)  }}$ and $\xi_{_{\left(  4\right)  }}$
have the phase part of the given gauge preferred.

\bigskip

\section{ACKNOWLEDGMENTS}

\bigskip

The author would like to thank Brian Edgar for drawing his attention to the
problem, for many useful discussions, and for a critical reading of the
manuscript. He is also grateful to the Natural Sciences and Engineering
Research Council of Canada for its ongoing financial support.

\end{document}